\documentclass[a4paper,11pt]{article}
\pdfoutput=1
\usepackage{jheppub}

\newcommand{\Za}{(Z\alpha)}
\newcommand{\e}{\epsilon}
\DeclareMathOperator*{\res}{Res}
\renewcommand{\Im}{\operatorname{Im}}

\title{Two-loop corrections to Lamb shift and hyperfine splitting in hydrogen via multi-loop methods.}

\author{Petr A. Krachkov}
\author{and Roman N. Lee}

\affiliation{Budker Institute of Nuclear Physics, Novosibirsk 630090, Russia}

\emailAdd{p.a.krachkov@inp.nsk.su}
\emailAdd{r.n.lee@inp.nsk.su}

\abstract{
	We revisit the contributions of order $\alpha^2(Z\alpha)^5m$ and $\alpha^2(Z\alpha)E_F$, respectively, to the Lamb shift and to the hyperfine splitting from mixed self-energy-vacuum-polarization diagrams, involving fermionic loop. We use modern multi-loop calculation techniques based on IBP reduction and differential equations. We construct the $\e$-regular basis \cite{Lee2019_eps} and explicitly demonstrate that it is compatible with the renormalization. We obtain analytic results in terms of one-fold integral involving elliptic function and dilogarithm. As a by-product, we obtain the analogous contribution for the limiting cases of heavy and light fermionic loop.
}

\begin{document}

\maketitle
\flushbottom

\section{Introduction}
Hydrogen atom is the simplest atomic system. Traditionally it served as a touchstone for testing the bound-state quantum electrodynamics (QED). At present, the precision of both theory and experiment for the electronic hydrogen has been increased to such an extent that comparison of calculated and measured transition energies can be used for the most accurate determination of the Rydberg constant once the contribution of proton (which is the hydrogen nucleus) structure is established. The latter gets much more pronounced in the (electronic) hydrogen cousin --- the muonic hydrogen. In particular, the Lamb shift in the muonic hydrogen provides the most precise value of the proton charge radius.\footnote{In that regard one should keep in mind the persisting controversy between the muonic hydrogen and electron-proton scattering experiments.}

The calculations of various contributions to the Lamb shift and the hypefine splitting have a long history starting from Refs. \cite{karplus1951electrodynamic,bethe1947electromagnetic,kroll1951radiative,karplus1952electrodynamic}, see also review \cite{Eides:2007exa} and references therein.
For higher order corrections these calculations often produced only numerical results, which might have insufficient accuracy for future comparison of theory and experiment. Also, despite these efforts, the present theoretical calculations are well behind the experimental measurements for some physical observables. E.g., the $1S-2S$ transition frequency measurements have reached accuracy of a few tens Hz, while the corresponding uncertainty of the available theoretical predictions is only a few tens of kHz. Therefore, new ways of Lamb shift calculations are very much welcome. Recent progress in multi-loop calculations provides an opportunity to apply the developed methods to such calculations.


\begin{figure}
	\centering
	\includegraphics[width=0.4\linewidth]{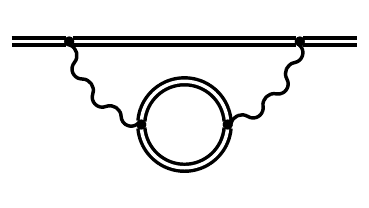}
	\caption{Two-loop self-energy diagram in Furry representation. Double lines denote electron propagators in the external electromagnetic field $A^{\mu}=\left(\,Z|e|/r,\  \boldsymbol{\mu}\times \boldsymbol{r}/r^3\, \right)$.}
	\label{fig:diagSE}
\end{figure}

In the present paper we apply the multi-loop methods to the calculation of the contribution to the Lamb shift (LS) and to the hyperfine splitting (HFS) of the diagram, depicted in Fig. \ref{fig:diagSE}. The double line denotes the electron propagator in the electromagnetic field of the nucleus $A^{\mu}=\left(\,Z|e|/r,\  \boldsymbol{\mu}\times \boldsymbol{r}/r^3\, \right)$. Here $Z|e|$ and $\boldsymbol{\mu}$ is the nucleus charge and magnetic moment, respectively. For LS calculations we neglect the nucleus magnetic field, while for the HFS calculations we consider linear in $\boldsymbol{\mu}$ contributions. Note that the magnetic moment of the nucleus is the vector operator ${\boldsymbol{\mu}}=\mu \boldsymbol{S}/S$. However, since we consider only linear in $\boldsymbol{\mu}$ contributions, we can formally treat it as a numeric vector up to the point where we average the result over the spin wave functions.

The leading in $Z\alpha$ contribution of the diagram in Fig. \ref{fig:diagSE} to LS and HFS is of the order $\mathcal{O}\left(\alpha^2\Za^4 m_e\right)$ and $\mathcal{O}\left(\alpha^2E_F\right)$, respectively.
This contribution is completely determined by the $\propto n_f$ terms\footnote{Here we follow the standard convention by introducing a formal parameter $n_f$ counting the number of electron loops.}  in the slope of the Dirac form factor and the value of the Pauli form factor of the electron at zero momentum transfer at two loops calculated long ago in Refs. \cite{Petermann:1957hs,Barbieri:1972as}. Therefore, we do not consider these contributions.
Here  $e$ and $m_e$ is the electron charge and mass, respectively,
\begin{equation}
	E_F=\frac{2S+1}{3S\pi}|e|\mu \Za^3 m_e^2
\end{equation} is the Fermi energy,\footnote{Substituting $Z=1,\,S=1/2, \mu =\frac{g}{2}\frac{|e|}{2M}$ for usual hydrogen nucleus, we obtain $E_F=\frac{8}{3}\frac{g}{2}\frac{m_e}{M}\alpha^4 m_e$.} $S$ is the nucleus spin, $\alpha=e^2/(4\pi)\approx1/137.036$ is the fine structure constant.

The next order contribution, $\mathcal{O}\left(\alpha^2 \Za^5 m_e\right)$ and $\mathcal{O}\left(\alpha^2 Z\alpha E_F\right)$, respectively, to LS and HFS comes from the diagrams, depicted in Figs. \ref{fig:diagLbL} and \ref{fig:diagFL}, which we will refer to as light-by-light (LbL) contribution and ``free loop'' (FL) contribution, respectively.

The light-by-light contributions to LS and HFS were calculated in  Refs. \cite{PhysRevA.48.2609,Eides1994a,Eides1993a,Eides1994} and \cite{Eides1991,Kinoshita1994}, respectively.
The result for the Lamb shift was presented in terms of a four-fold integral, while that for the hyperfine splitting was presented in terms of a three-fold integral. These integrals were then treated numerically. The free loop contribution to Lamb shift was obtained in terms of two-fold integrals in Refs. \cite{PhysRevA.48.2609,Eides1993}, while the corresponding contribution to the hyperfine splitting was obtained in Refs. \cite{Eides1990,Kinoshita1994} in terms of one-fold integral involving elliptic function and logarithm (or rather, arctangent).

In a sense, the result of the present paper is the representation of all four contributions (LbL and FL to LS and HFS) in the form similar to that of Ref. \cite{Eides1990}. We use modern multi-loop methods, namely, the IBP reduction \cite{Chetyrkin1981} and the differential equations method \cite{Kotikov1991, Remiddi1997}. In order to apply the latter, we cosider the diagrams where the mass of the fermion in the loop is different from that of the electron line. As a by-product, we also obtain the contribution of the muon loop for the ordinary hydrogen and that of electron loop for the muonic hydrogen.

\section{Calculation}
\begin{figure}
	\centering
	\includegraphics[width=0.6\linewidth]{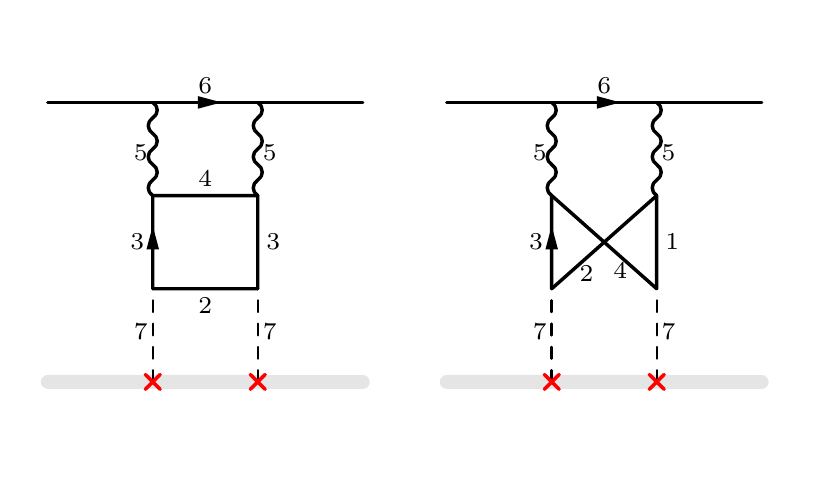}
	\caption{Gauge invariant set of diagrams which corresponds to the light-by-light contribution to the Lamb shift. Numbers correspond to the enumeration of denominators in Eq. \eqref{eq:Ds_LbL}.}
	\label{fig:diagLbL}
\end{figure}

\begin{figure}
	\centering
	\includegraphics[width=0.8\linewidth]{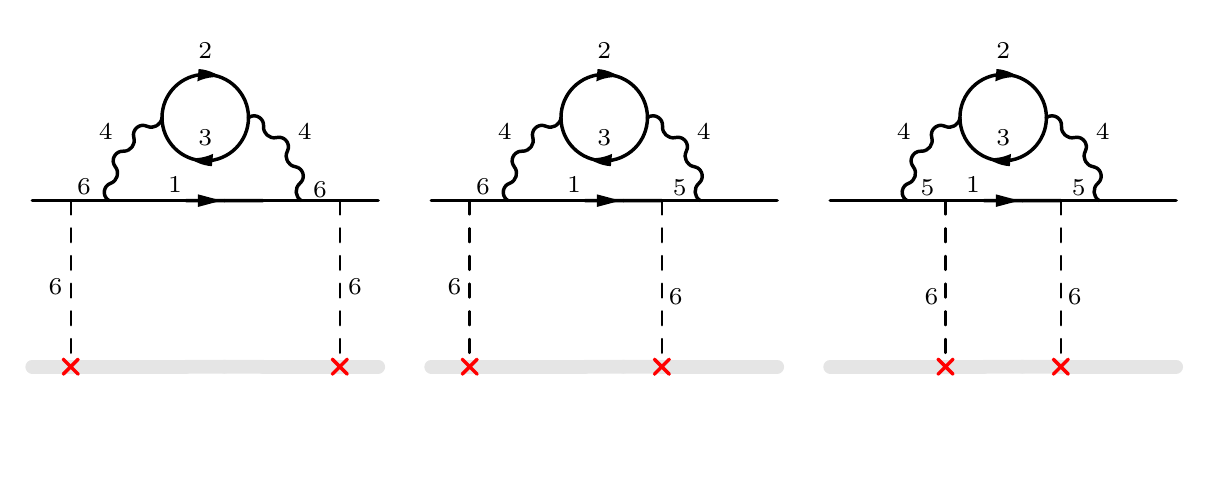}
		\includegraphics[width=0.8\linewidth]{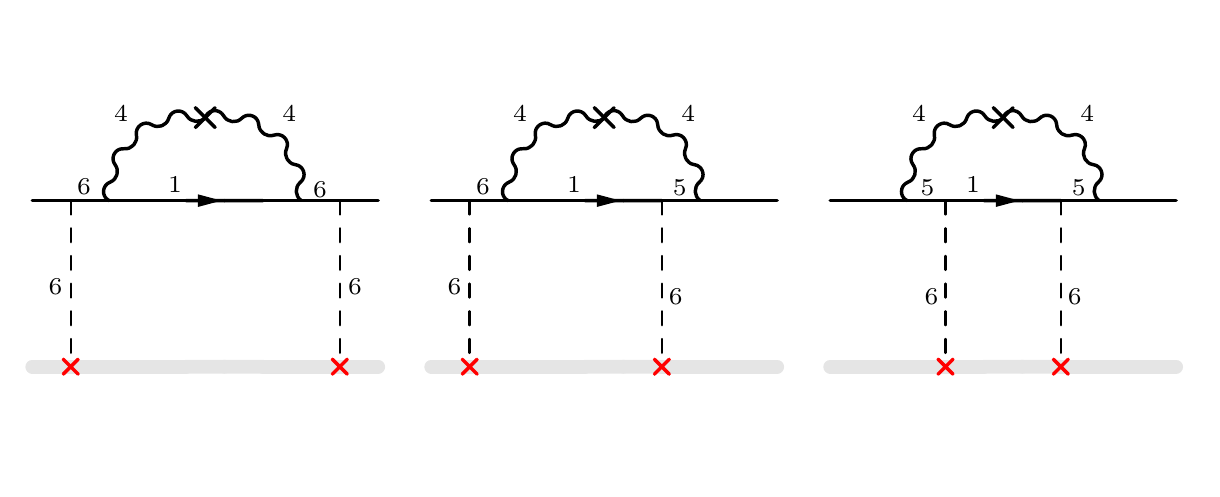}
	\caption{Gauge invariant set of diagrams which corresponds to the free-loop contribution to the Lamb shift. The second line of diagrams is a one-loop counter-terms. Numbers correspond to the enumeration of denominators in Eq. \eqref{eq:Ds_FL}.}
	\label{fig:diagFL}
\end{figure}

We consider the diagrams depicted in Figs. \ref{fig:diagLbL} and \ref{fig:diagFL}.
The contribution of each set is gauge invariant and can be calculated independently. Note that for the HFS calculations one should replace one of the two Coulomb exchanges in those diagrams with the magnetic exchange corresponding to the contribution of the nuclear magnetic moment.

The external electron legs on the diagrams denote the bound electron wave function, however, with the precision that we pursue here, the bound-state effect is properly taken into account by the factor $|\psi_{n,l}(0)|^2= \delta_{l,0}\frac{(m_e Z\alpha)^3}{\pi n^3}$, where $\delta$ is a Kronecker symbol, $n$ and $l$ are a principal and azimuthal quantum number, correspondingly. Indeed,  the characteristic loop momenta in the discussed diagrams are  $k_i\sim m_e$, while that of the wave function $|\boldsymbol{p}|\sim m_e Z\alpha\ll m_e$. Therefore, we calculate the diagrams in Figs. \ref{fig:diagLbL} and \ref{fig:diagFL} for the external electron legs corresponding to free electron with momentum $p=m_e \tau$, where $\tau=(1,\boldsymbol{0})$ is the time ort.

The LbL contribution is both UV and IR finite, while the FL contribution is UV and IR divergent. In order to obtain the finite result we add one-loop counter terms, depicted in  Fig. \ref{fig:diagFL} (second line). The two-loop counter terms, as well as the IR subtraction are expressed via scaleless integrals which are equal to zero in dimensional regularization. In particular, denoting the momentum of the Coulomb line as $k_1=(0,\boldsymbol{k}_1)$, we see that the two-loop counter terms contain only one scaleless denominator $(p-k_1)^2-m_e^2=k_1^2$.

The Lamb shift energy correction can be represented as
\begin{equation}\label{eq:Lamb:start}
\delta E_{\text{LS}}=\frac{1}{2 m_e}\overline{u}\hat O u|\psi_{n,l}(0)|^2=\frac{1}{4 m_e}\operatorname{Tr}\left[\hat O (\hat p+m_e)\right]|\psi_{n,l}(0)|^2\,,
\end{equation}
where $\overline{u}\hat O u$ corresponds to the diagrams in Figs.  \ref{fig:diagLbL} and \ref{fig:diagFL} with both incoming and outgoing momenta equal to $p=m_e \tau$.

In the case of hyperfine splitting we should replace one Coulomb exchange with the magnetic exchange. Since we use the dimensional regularization, we should avoid using $\gamma^5$ and $\varepsilon_{ijk}$ because those objects are poorly generalized to the generic spacetime dimension $d=4-2\e$. Therefore we rewrite all formulas involving vector product $(\boldsymbol{a}\times \boldsymbol{b})_k=\varepsilon_{ijk}a_ib_j$ in terms of antisymmetric tensors $a_ib_j-a_jb_i$. We obtain
\begin{equation}\label{eq:Hyp:start}
\delta E_{\text{HFS}}=\frac{1}{24 m_e}\operatorname{Tr}\left[\hat O^{i,j}_{\text{HFS}} (\gamma_i\gamma_j-\gamma_j\gamma_i)(\hat p+m_e)\right]|\psi_{n,l}(0)|^2\,,
\end{equation}
where
$\overline{u}\hat O^{i,j}_{\text{HFS}}u$ corresponds to the sum of the diagrams in Figs.  \ref{fig:diagLbL} and \ref{fig:diagFL} in which one of the two Coulomb exchanges $Z|e|\gamma^0/\boldsymbol{k}_1^2$ is replaced by the ``magnetic exchange'' $\mu{\gamma^ik_1^j}/{\boldsymbol{k}_1^2}$.

\subsection*{Differential system and boundary conditions.}

In order to apply the differential equations method, we decouple the mass of the bound electron and the mass of the particle in the loop. We put the latter to $1$ while keeping the mass of the bound electron as a free parameter $m$.

For the light-by-light contribution we consider the integral family
\begin{equation}
\label{eq:j_LbL}
j_{\text{LbL}}(n_{1},\ldots,n_{9})=\int\frac{dk_1dk_2 dk_3}{\pi^{3d/2}}
\prod_{k=1}^{8}\left[D_{k}+i0\right]^{-n_{k}}\times
\frac{\delta^{(n_9-1)}\left(-D_{9}\right)}{(n_9-1)!}\,,
\end{equation}
where
\begin{align}\label{eq:Ds_LbL}
&D_1=(k_1-k_2+k_3)^2-1\,,\quad D_2=(k_1-k_2)^2-1\,,\quad D_3=k_2^2-1\,,\nonumber\\
&D_4=(k_2-k_3)^2-1\,,\quad D_5=k_3^2\,,\quad D_6=(k_3+p)^2-m^2\,,\quad D_7=k_1^2\nonumber\\
& D_8=2k_2\cdot p\,,\quad D_9=2 k_1\cdot \tau =2 k_1^{0}\,.
\end{align}
The $\delta$-function in Eq.\eqref{eq:j_LbL} corresponds to zero energy transfer on the nucleus. Note that $n_8$ is not positive.

For the free-loop contribution we consider the family
\begin{equation}\label{eq:j_FL}
 j_{\text{FL}}(n_{1},\ldots,n_{9})=\int\frac{dk_1dk_2 dk_3}{\pi^{3d/2}}
\prod_{k=1}^{8}\left[\tilde D_{k}+i0\right]^{-n_{k}}\times
\frac{\delta^{(n_9-1)}\left(-\tilde D_{9}\right)}{(n_9-1)!}\,,
\end{equation}
where
\begin{align}\label{eq:Ds_FL}
&\tilde D_1=(k_1-k_3)^2-m^2\,,\quad \tilde D_2=k_2^2-1\,,\quad \tilde D_3=(k_2-k_3)^2-1\,,\nonumber\\
&\tilde D_4=(k_3)^2\,,\quad \tilde D_5=(p-k_3)^2-m^2\,,\quad \tilde D_6=k_1^2\,,\quad \tilde D_7(k_1-k_2)^2-1\nonumber\\
&\tilde D_8=2k_2\cdot p\,,\quad \tilde D_9=D_9=2 k_1\cdot \tau=2 k_1^0\,,
\end{align}
where $n_7$ and $n_8$ are not positive.

\begin{figure}[h]
	\centering
	\includegraphics[width=1.\linewidth]{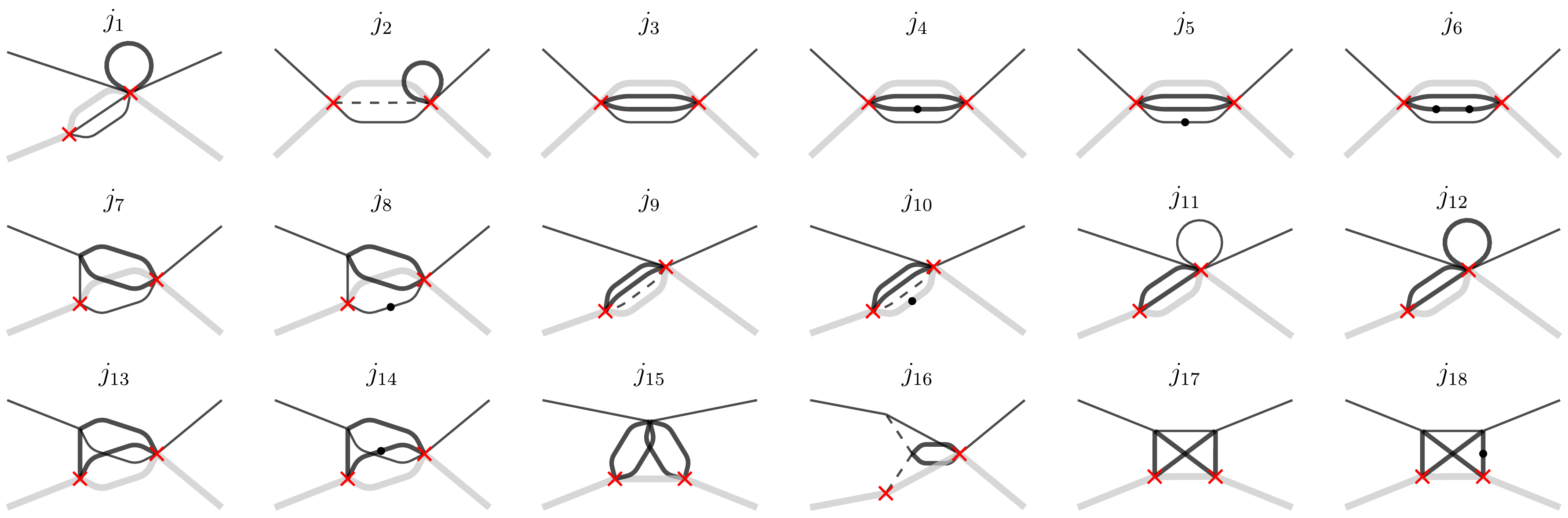}
	\caption{Basis of master integrals. Thick and thin black lines correspond to the denominators with mass $1$ and $m$, respectively. Thick gray line corresponds to $\delta(-D_9)$, dashed lines correspond to  massless denominators.}
	\label{pic_masters}
\end{figure}

Making the IBP reduction \cite{Chetyrkin1981,Tkachov1981} with \texttt{LiteRed} \cite{Lee2014}, we reveal $14$ master integrals for  light-by-light contribution and $8$ master integrals for free-loop contribution. Four master integrals are common for the two bases. Thus in the merged basis we have $14+8-4=18$ master integrals presented in Fig. \ref{pic_masters}. The first eight graphs correspond to FL basis, while the graphs $3-6$ and $9-18$ correspond to LbL basis\footnote{Note that the master integral $j_{10}$, according to Eq. \eqref{eq:j_LbL}, contains the derivative of $\delta$-function which is odd with respect to the substitution $k_1\to -k_1$. The remaining factor is easily shown to be an even function, therefore, $j_{10}$ is identically zero.}.

Using the IBP reduction, we construct the differential equations  \cite{Kotikov1991,Remiddi1997} for the master integrals with respect to $m^2$:
\begin{equation}\label{eq:de_m}
\partial_{m^2} \boldsymbol j= M \boldsymbol j\,,
\end{equation}
where $\boldsymbol j = (j_1,...,j_{18})^\intercal$ is a  column of master integrals.

We use \texttt{Libra} \cite{Lee2021} to manipulate the differential system \eqref{eq:de_m}. We find it convenient to work with the master integrals in $2-2\e$ dimensions and later express the master integrals in $4-2\e$ dimensions via lowering dimensional recurrence relation \cite{Tarasov1996}. Note that the differential system \eqref{eq:de_m} can not be reduced to $\e$-form due to integrals $j_{3-6}$ which appear both in FL and LbL contributions. Those four master integrals can be expressed via hypergeometric functions. In particular,
\begin{multline}\label{eq:j3}
	j_3^{(2-2\e)}=\tfrac{2^{1-8 \e } m^{-2 \e} \Gamma (4 \e )}{\Gamma (\e +1)}\, _4F_3\left(\tfrac{1}{4}-\tfrac{\e }{2},\tfrac{3}{4}-\tfrac{\e }{2},\e +\tfrac{1}{2},2 \e +\tfrac{1}{2};\tfrac{1}{2}-\e ,1-\e ,\e +1;m^2\right)
	\\
	-\tfrac{2^{-1-6 \e} \Gamma \left(2 \e +\tfrac{1}{2}\right) \Gamma \left(3 \e +\tfrac{1}{2}\right) }{\sin (\pi  \e )  \Gamma (\e +1)^2}\, _4F_3\left(\tfrac{\e }{2}+\tfrac{1}{4},\tfrac{\e }{2}+\tfrac{3}{4},2 \e +\tfrac{1}{2},3 \e +\tfrac{1}{2};\tfrac{1}{2},\e +1,2 \e +1;m^2\right)\,.
\end{multline}
Due to this reason the corresponding block of matrix $M$ in Eq. \eqref{eq:de_m} can not be reduced in $\e$-form. All other blocks can be reduced to $\e$-form. With some trial and error method we have succeeded to obtain the $(A+\e B)$-form of the differential equation,
\begin{equation}
	\partial_{m^2} \boldsymbol J=M_1 \boldsymbol J\,,
\end{equation}
where $M_1(m^2,\e)=A(m^2)+\e B(m^2)$ and the matrix $A$ has nonzero entries  only in the columns $3$--$6$. We fix the boundary conditions at the point $m^2=0$.  The general solution has the form
\begin{equation}
	\boldsymbol J(m^2) =U(M_1|m^2,\underline{0})\boldsymbol{C}^{0}\,,
\end{equation}
where $\boldsymbol{C}^{0}$ is a column of the boundary constants and
\begin{equation}\label{eq:gen_sol}
	U(M|y,\underline{x})=\lim_{x_0\to x} \operatorname{Pexp}\left[\intop_{x_0}^{y}d\xi M(\xi,\e) \right]x_0^{\res_{\xi=x} M(\xi,\e)}.
\end{equation}
Using \texttt{Libra}, we relate the boundary constants $\boldsymbol{C}^{0}$ to specific asymptotic coefficients of original master integrals $j_{1-18}$ at $m^2\to 0$. We compute these coefficients by using expansion-by-regions method \cite{Beneke:1997zp} and direct integration of Feynman parametrization.
We find that the boundary constants $\boldsymbol{C}^{0}$ are expressed in terms of the following 5 nonzero constants:
\begin{align*}
	\left[j_1\right]_{m^{-1-4\e}}
	&=\left[j_{11}\right]_{m^{-2\e}}
	=\left[j_{12}\right]_{m^{0}}=\frac{ \Gamma \left(\frac{1}{2}+\e\right) \Gamma \left(\frac{1}{2}+2 \e \right)}{2^{1+2 \e}\pi  \e}\,,\
	\left[j_2\right]_{m^{-1-4\e}}=\frac{\Gamma \left( -\frac{1}{2}-3 \e\right) \Gamma (4 \e )}{2^{-1+6 \e } \sqrt{\pi } \Gamma (-2 \e )}\,,\\
	\left[j_3\right]_{m^{0}}&=\left[j_9\right]_{m^{0}}=\frac{\Gamma \left(\frac{1}{2}+2 \e\right) \Gamma \left(\frac{1}{2} + 3 \e \right) \Gamma (-\e )}{2^{1+6 \e} \pi  \Gamma (1+\e)}\,,\
	\left[j_3\right]_{m^{- 2\e}}=\frac{2^{1-8 \e } \Gamma (4 \e )}{\Gamma (1+\e)}\,,
\end{align*}
\begin{multline}
	\left[j_{15}\right]_{m^0}=j^{(2-2\e)}_{15}=\frac{2^{1+2\e} \Gamma (-4 \e ) \Gamma \left(\frac{1}{2}+2 \e\right) \Gamma \left(\frac{3}{2}+3 \e \right)}{4 (1 + 2 \e) \Gamma \left(\frac{1}{2}-\e \right)^2\tan (\pi  \e ) }
	\\
	+\frac{\Gamma \left(\frac{1}{2}+2 \e\right) \Gamma \left(\frac{1}{2}+\e\right)}{4^{1+\e}\pi  }\bigg[\, _3F_2\left(1,-\e ,\e +1;\tfrac{3}{2},\tfrac{1}{2}-2 \e ;1\right)\\
	-\frac{4 \e +1}{4 \e  (\e +1)}\, _3F_2\left(\tfrac{1}{2},1,2 \e +\tfrac{3}{2};1-\e ,\e +2;1\right)\bigg]\,,
\end{multline}
where $[j_k]_{m^\nu}$ denotes the coefficient in front of ${m^\nu}$ in the small-mass asymptotics of $j_k^{(2-2\e)}$.

All but the last constant are trivially expressed in terms of alternating multiple zeta values. It is not obvious which transcendental numbers might enter the expansion of the last constant $j_{15}^{(2-2\e)}$. Nevertheless, using the PSLQ and our experience, we have been able to establish that the $\e$-expansion of $j_{15}^{(2-2\e)}$ can be written in terms of Goncharov's polylogarithms at fourth root of unity. The two first terms of $\e$-expansion read
\begin{equation}
	j_{15}^{(2-2\e)}=e^{-3\e\gamma_E}\left[\tfrac{1}{2}-\tfrac{\pi }{8}+ \left(2G+\tfrac{3}{2} \pi  \ln2-4 \ln2-\tfrac{\pi }{2}-1\right) \e +\mathcal{O}\left(\e ^2\right)\right]\,,
\end{equation}
where $G = \Im{\text{Li}_2(i)} = \sum _{k=0}^{\infty } \frac{(-1)^k}{(2 k+1)^2}= 0.915965594...$ is the Catalan constant. For our present goal we will need only the leading term of this expansion.

\subsection*{$\e$-regular basis}

Since the differential system \eqref{eq:de_m} for the master integrals can not be reduced to $\e$-form, we should rely on the Frobenius method for the calculation of the evolution operator $U(m^2,\underline{0})$ in Eq. \eqref{eq:gen_sol}. The dependence of this operator on $\e$ constitutes substantial technical difficulties and blocks the way to high-precision numerical calculation suitable for the application of PSLQ algorithm, \cite{ferguson1999analysis}, for the recognition of the master integrals in the point $m=1$. Thus we choose to switch to the $\e$-regular basis \cite{Lee2019_eps}. After finding this basis, we simply put $\e=0$.

Note that the counter-term diagrams in the second line of Fig. \ref{fig:diagFL} can also be expressed in terms of the three-loop master integrals in Fig. \ref{pic_masters} although they have only two loops. To this end we multiply the corresponding integrals by $1=\frac{-1}{\Gamma[1-d/2]}\int \frac{d^d k_2}{i\pi^{d/2}D_2}$. Then the contribution of counter-terms is expressed via the master integrals with unit mass tadpole loop, namely, via $j_1,\ j_2$. Luckily, $\Gamma[1-d/2]$ in the denominators cancels with the same $\Gamma$ in $\delta Z_3^{(1l)} = -\frac{4\Gamma(2-d/2)\alpha}{3(4\pi)^{d/2-1}}= -\frac{4(1-d/2)\Gamma(1-d/2)\alpha}{3(4\pi)^{d/2-1}}$ which stands for the cross in the couter-term diagrams. Therefore, the finite sum of all diagrams in Fig. \ref{fig:diagFL} is expressed via the integrals of the family \eqref{eq:j_FL} with \textbf{rational} coefficients\footnote{By ``rational'' we understand the coefficients which are rational as functions of both $\e$ and $m^2$.}. Then we are in position to state the existence of $\e$-regular basis \cite{Lee2019_eps}.

Let us describe how we construct this $\e$-regular basis.
\begin{enumerate}
	\item We start from the found $(A+\e B)$-form. Thus, the coefficients of differential equations are regular in the limit $\e\to0$.
	\item We determine the highest leading term $\mathcal{O}(\e^{-n})$ in $\e$-expansion of the boundary constants. We multiply \textbf{all} integrals by $\e^n$ to make all new functions $\boldsymbol{J}$ finite. The two next steps are performed in the loop over the row number $i$.
	\item Then we use the following rule of thumb: if for, say, $J_i$ function both the boundary constant $C_i$ and the right-hand side of the equation are zero at $\e=0$  is zero, then the function itself is also zero at $\e=0$. Therefore, we redefine  $J_i\to \e J_i$ and modify respectively the boundary constant $C_i\to \e C_i$ and the differential system. The latter modification is reduced to the multiplication/division by $\e$ the $i$-th column/row of the matrix $A+\e B$.
	\item We also use substitutions of the form $J_i\to J_i+\sum_{k<i} c_k J_k$, where the coefficients $c_k$ are \textbf{rational numbers} chosen so as to nullify as many entries on the $i$-th row of the matrix $M(m^2,0)$ as possible.
\end{enumerate}

This approach works perfectly for all rows except for the rows $3-6$ corresponding to the equations irreducible to $\e$-form. For those $4$ master integrals we use the presumable finiteness of the sum of diagrams in Figs. \ref{fig:diagLbL},\ref{fig:diagFL} as a guiding line for finding the relations between $j_{3-6}$ near $d=2$. We find that
\begin{multline}
	Q(m^2,\e)=\left(1-m^2\right) \left(2 + m^2 \e +12 \e\right) j_3
	+\left(78 m^4 \e -25 m^4-72 m^2 \e +36 m^2-8\right)j_4\\
	-m^2 \left(4 -m^2+6 m^2 \e\right)j_5
	-48 \left(1-m^2\right) m^2 (1-4 \e) j_6 =\mathcal{O}(\e^2)
\end{multline}
Thus we choose $J(m^2,\e)=Q(m^2,\e)/\e^2$ as an element of $\e$-regular basis. Indeed, we see that after this choice both the boundary constants and the matrix $M(m^2,\e)$ have finite limit $\e\to 0$.

The result of our approach is the differential system in which we can simply put $\e=0$. The higher orders in $\e$ which we miss with putting $\e$ to zero are guaranteed not to appear in our final results, which is the rationale behind the notion of $\e$-regular basis. The resulting system has the form
\begin{equation}
\partial_{m^2} \boldsymbol J_{\text{reg}}= \overline{M}(m^2) \boldsymbol J_{\text{reg}}\,,
\end{equation}
where the matrix $\overline M$ is strictly lower triangular except for the diagonal $4\times4$ block with indices $3$--$6$. It is essential that $\overline M$ does not depend on $\e$. The singular points of the differential system are $m^2=0,1,\infty$. Again, we write the solution in the form
\begin{equation}\label{eq:dsreg}
\boldsymbol J_{\text{reg}}(m^2)=U(\overline{M}|m^2,\underline{0}) \boldsymbol C^{0}\,.
\end{equation}
The almost lower triangular structure of the matrix $M$ allows us to write the general solution in terms of polylogarithms and/or one-fold integrals of $j_{3-6}$ multiplied by polylogarithms. However we choose here to construct the Frobenius expansions near each of the three singular points of the differential system \eqref{eq:dsreg}, $m^2=0,1,\infty$. We match the obtained expansions pairwise in the points which belong to the intersection of convergence regions of the corresponding two expansions. For example the expansions near $m^2=0$ and  $m^2=1$ are connected via the relation
\begin{equation}
\boldsymbol J_{\text{reg}}(1/2)=U(\overline{M}|1/2,\underline{0}) \boldsymbol C^0=U(\overline{M}|1/2,\underline{1}) \boldsymbol C^1\,.
\end{equation}
Then the boundary constants at $m^2=1$ are expressed as $\boldsymbol{C}^1=U^{-1}(\overline{M}|\tfrac12,\underline{1})U(\overline{M}|\tfrac12,\underline{0})\boldsymbol{C}^0$. We calculate $1000$ terms of series of $U(1/2,\underline1)$ and $U(1/2,\underline0)$ and compute more than $300$ digits for $\boldsymbol{C}^{1}$. In order to use PSLQ recognition, we need to have a basis of appropriate transcendental numbers and we extract all but one required nontrivial constants from Ref. \cite{Eides1990}. Therein the result for the FL contribution to HFS was expressed in terms of the weighted sum of the integrals (cf. Eq. (11) of Ref. \cite{Eides1990})
\begin{equation}\label{eq:cr13}
	\{c_1,c_2,c_3,c_4\}=\intop_0^1\frac{dq}{q^2}[\mathrm{K}(q^2)-\mathrm{E}(q^2)]\left\{\frac{\arctan\left(\sqrt{\frac{2 q}{1-q}}\right)}{1+q},\,\frac{\sqrt{\frac{2 q}{1-q}}}{1+q},\,\sqrt{\frac{2 q}{1-q}},\,q\sqrt{\frac{2 q}{1-q}} \right\}\,,
\end{equation}
where $\mathrm{K}(x)$ and $\mathrm{E}(x)$ are complete elliptic integrals.
Moreover, using PSLQ it is easy to establish a relation $6c_2-5c_3+2c_4=0$ overlooked in Ref. \cite{Eides1990}. Using some guess work, we find the following basis sufficient for our purpose
\begin{align}\label{eq:constants}
\boldsymbol B&= \{1,\pi,\ln 2,\pi^2,\zeta_3,e_0,1/e_0,e_1,e_2\}\,,\nonumber\\
e_0&=\frac{3}{4\pi}\intop_0^1  \frac{dq\,\mathrm{K}(q^2)}{1+q}\sqrt{\frac{2q}{1-q}}=\tfrac{\Gamma(3/8)\Gamma(9/8)}{\Gamma(5/8)\Gamma(7/8)}=1.42812528609616838918477155113\ldots\,,\nonumber\\
e_1&=\frac{2}{\pi}\intop_0^1  \frac{dq\,\mathrm{K}(q^2)}{1+q}\arctan\left(\sqrt{\frac{2q}{1-q}}\right)=0.70733097502159315134278673801\ldots\,,\nonumber\\
e_2&=\frac{2}{\pi}\intop_0^1 \frac{dq \,\mathrm{K}(q^2)}{1+q}\Im\mathrm{Li}_2\left(i\sqrt{\frac{2q}{1-q}}\right) =1.08354966910460443406693681278\ldots\,,
\end{align}
where $\mathrm{Li}_2(x)$ is a dilogarithm. The constants $c_{1-4}$ in Eq. \eqref{eq:cr13} are expressed as linear combinations with rational coefficients of $ \{1,e_0,1/e_0,e_1\}$:
\begin{equation}
	c_1=\tfrac{1}{2 e_0}-\tfrac{7 e_0}{6}+e_1+1, \quad
	c_2=\tfrac{1}{e_0}+\tfrac{e_0}{3}, \quad
	c_3=\tfrac{1}{e_0}+e_0, \quad
	c_4=\tfrac{3 e_0}{2}-\tfrac{1}{2 e_0}\,.
\end{equation}
The benefit of using $e_0,e_0^{-1},$ and especially $e_1$ instead of $c_{1-4}$ is that their form allowed us to guess the form of the last nonstandard constant $e_2$, which was deduced from $e_1$ by noting that $\arctan\left(x\right)=\Im\mathrm{Li}_1(i x)$ and then replacing $\mathrm{Li}_1\to \mathrm{Li}_2$. To obtain the contributions to the Lamb shift and the hyperfine splitting we need a few coefficients of the expansion of $J_{\text{reg}}(m^2)$ near the point $m^2=1$. This expansion has the form
\begin{equation}\label{eq:series_at_1}
	\boldsymbol J_{\text{reg}}(m^2)=U(\overline{M}|m^2,\underline{1}) \boldsymbol C^1\,,
\end{equation}
where $U(\overline{M}|m^2,\underline{1})$ has the form of generalized power series in $1-m^2$. In particular, $U(m^2,\underline{1})$ contains $\ln (1-m^2)$. However, we have checked that these logarithms disappear when $U(m^2,\underline{1})$ is multiplied by $\boldsymbol C^1$ so that the specific solution has no branching at the point $m^2=1$ as it should be.

The power series \eqref{eq:series_at_1} converge when $m^2\in(0,2)$.
To obtain the results for $\boldsymbol J_{\text{reg}}(m^2)$ at $m^2>2$ we pass to the variable $z=\frac{m^2-1}{m^2}$ and again match the power series near $z=0$ ($m^2=1$) and $z=1$ ($m^2=\infty$) at $z=1/2$. In this way we obtain the high-precision numerical result for the column of boundary constants $\boldsymbol{C}^{\infty}$ which define the coefficients in the asymptotic expansion of $\boldsymbol J_{\text{reg}}(m^2)$ at $m^2\to\infty$. In order to define the analytic form of $\boldsymbol{C}^{\infty}$ we again use PSLQ recognition with the following basis:
\begin{equation}
\boldsymbol B_\infty= (1,\pi,\ln 2,\pi^2,\ln^2 2,\zeta_3,i_0,1/i_0)\,,\qquad
i_0=\tfrac{\Gamma(5/4)^2}{\Gamma(3/4)^2}\,.
\end{equation}
The nontrivial constants $i_0$ and $i_0^{-1}$ were conjectured by examining the large-mass asymptotics of $j_3$ from Eq. \eqref{eq:j3}.

\section{Results}
The discussed corrections to the Lamb shift and hyperfine splitting have the following form
\begin{align}\label{eq:res0}
\delta E^{i}_{\text{LS}}&=\delta_{l,0} \frac{\alpha^2(Z\alpha)^5}{\pi n^3}  m_e  \left(\frac{m_r}{m_e}\right)^3 B_{i}\,,\nonumber\\
\delta E^{i}_{\text{HFS}}&=\delta_{l,0} \frac{\alpha^2(Z\alpha)}{\pi n^3} E_F \left(\frac{m_r}{m_e}\right)^3 C_{i}\,,
\end{align}
where $i=FL,LbL$ for different contributions and we have recovered the recoil factor $\left(\frac{m_r}{m_e}\right)^3=\left(\frac{M}{M+m_e}\right)^3$ accounting the main effect of the finite nuclear mass $M$.

For the case of electron loop in electron atom we have the following result:
\begin{align}
B_{FL}&=
-\frac{17267}{7560}-\frac{241}{324 e_0}+\frac{296767 e_0}{132300}-\frac{2e_1}{3}
=-0.072909964469261993898\ldots
\,,\nonumber\\
B_{LbL}&=
\frac{21401}{540}-\frac{383}{54 e_0}-\frac{11 e_0}{10}+\frac{10e_1}{3}+\frac{16e_2}{9}+\frac{13\pi}{36}-\frac{\pi^2}{12}-\frac{406}{9} \ln {2}-\frac{49}{9} \zeta_3
\nonumber\\
&\mspace{303mu}=-0.12291622969679641051\ldots
\,,\nonumber\\
C_{FL}&=
-\frac{59}{270}-\frac{343}{1296 e_0}-\frac{1079 e_0}{10800}+\frac{e_1}{3}
=-0.31074204276601754458\ldots
\,,\nonumber\\
C_{LbL}&=
46-\frac{39}{2 e_0}-\frac{23 e_0}{6}+16 e_1+\frac{40 e_2}{9}-\frac{\pi^2}{2}-32 \ln {2}-\frac{245}{18} \zeta_3
\nonumber\\
&\mspace{267mu}=-0.47251462820471059187\ldots
\,.
\end{align}
Given the representation of Eq. \eqref{eq:constants} for the constants $e_0$, $e_1$, $e_2$, the above expressions provide a one-fold integral representation for the corresponding contributions to Lamb shift and hyperfine splitting. The numerical values for $B_{FL}$, $B_{LbL}$,  $C_{FL}$,  $C_{LbL}$ agree with the results of Refs. \cite{PhysRevA.48.2609,Eides1993,Eides1994a,Eides1994,Eides1990,Eides1993a,Kinoshita1994}.
For the case of muon loop in electron atom we have the following result:
\begin{align}
B_{FL}&=
-\frac{7}{60}\left(\tfrac{m_e}{m_\mu}\right)+\frac{109}{1512}\left(\tfrac{m_e}{m_\mu}\right)^2-\frac{3}{56}\left(\tfrac{m_e}{m_\mu}\right)^3
+\mathcal{O}\left(\left(\tfrac{m_e}{m_\mu}\right)^4\right)\,,\nonumber\\
B_{LbL}&=
\left[\frac{470561}{21600}-\frac{437 \pi}{720}-\frac{27647}{960}\ln{2}  +\frac{281}{2304} \ln\left(\tfrac{m_e}{m_\mu}\right)\right] \left(\tfrac{m_e}{m_\mu}\right)^2
+\mathcal{O}\left(\left(\tfrac{m_e}{m_\mu}\right)^4\right)\,,\nonumber\\
C_{FL}&=
-\frac{1}{3}\left(\tfrac{m_e}{m_\mu}\right)+\frac{1751}{30240} \left(\tfrac{m_e}{m_\mu}\right)^3
+\mathcal{O}\left(\left(\tfrac{m_e}{m_\mu}\right)^4\right)\,,\nonumber\\
C_{LbL}&=
\left[\frac{5}{24}-2\pi +8 \ln{2}\right]\left(\tfrac{m_e}{m_\mu}\right)
\nonumber\\
&+\left[\frac{6931861}{552960}-\frac{\pi}{3}- \frac{38107}{2304}\ln{2} -\frac{805}{9216} \ln\left(\tfrac{m_e}{m_\mu}\right)\right] \left(\tfrac{m_e}{m_\mu}\right)^3
+\mathcal{O}\left(\left(\tfrac{m_e}{m_\mu}\right)^4\right)\,.
\end{align}

One is tempted to obtain also the results for the contribution of the electron loop in the muonic hydrogen by making a substitution  $m_e\to m_\mu$ in Eq.~\eqref{eq:res0}  and using the large-$m$ asymptotics of $\boldsymbol{J}_{\text{reg}}$. Then the results for functions $B_{i}$ and $C_{i}$  have the following form
\begin{align}\label{eq:large}
B_{FL}&=
-i_0\frac{1600}{441}\sqrt{\tfrac{m_e}{m_\mu}}
+\mathcal{O}\left(\tfrac{m_e}{m_\mu}\right)\,,\\
B_{LbL}&=
\left(-\frac{139}{18}+\frac{2 \pi^2}{3}+\frac{4}{3}\ln{2}\right)\tfrac{m_\mu}{m_e} + \frac{4187}{135}+\frac{7 \pi}{36}-\frac{406}{9}\ln{2}
+\mathcal{O}\left(\tfrac{m_e}{m_\mu}\right)\,,\nonumber\\
C_{FL}&=
\left(-\frac{13}{6}+\frac23\ln{2}\right)\ln\left(\tfrac{m_\mu}{m_e}\right) +\frac{379}{72}-\frac{14}{9}\ln{2}-\frac{\pi^2}{18}+ \frac{1}{3}\ln^2{2}-\frac{72}{25 i_0}\sqrt{\tfrac{m_e}{m_\mu}}
+\mathcal{O}\left(\tfrac{m_e}{m_\mu}\right)\,,\nonumber\\
C_{LbL}&=
\left(12+4\ln{2}-\frac{5\pi^2}{3}\right)\ln\left(\tfrac{m_\mu}{m_e}\right)-13+\frac{5\pi^2}{6} +2\ln^2{2}+5\zeta_3
+\mathcal{O}\left(\tfrac{m_e}{m_\mu}\right)\,.\nonumber
\end{align}

However, for muonic atom the characteristic momenta of bound muon (or the inverse size of its wave funcetion), $m_\mu Z\alpha$ is of the same order, or larger, as the characteristic loop momenta $m_e$ even for $Z=1$. Therefore, the applicability condition for the approximation used in the present paper is violated and Eq. \eqref{eq:large} may be considered at most as an order of magnitude estimate.

\section{Conclusion}

In the present paper we revisit the  contributions of order $\alpha^2(Z\alpha)^5m$ to the Lamb shift and of the order $\alpha^2(Z\alpha)E_F$ to the hyperfine splitting from mixed self-energy-vacuum-polarization diagrams depicted in Figs. \ref{fig:diagLbL} and \ref{fig:diagFL}. We construct the $\e$-regular basis \cite{Lee2019_eps} and explicitly demonstrate that its elements taken at $\e=0$ are sufficient to express the renormalized results. The results have the form of one-fold integral involving elliptic function and dilogarithm and agree with previously known numerical results. We also obtain the contribution of the same set of diagrams with different mass of the fermion in the loop and in the fermion line, which allows us to determined the corresponding contribution of the muonic loop in the conventional hydrogen as well as the estimate of the electron loop contribution in the muonic hydrogen.

\acknowledgments
The work has been supported by Russian Science Foundation under grant 20-12-00205.


\providecommand{\href}[2]{#2}\begingroup\raggedright\endgroup
\end{document}